\documentclass[conference]{IEEEtran}
\IEEEoverridecommandlockouts
\usepackage{cite}
\usepackage{amsmath,amssymb,amsfonts}
\usepackage{algorithmic}
\usepackage{graphicx}
\usepackage{textcomp}
\usepackage{xcolor}
\def\BibTeX{{\rm B\kern-.05em{\sc i\kern-.025em b}\kern-.08em
    T\kern-.1667em\lower.7ex\hbox{E}\kern-.125emX}}
\begin{document}


%

\newcommand*{\Scale}[2][4]{\scalebox{#1}{$#2$}}%
\newcommand*{\Resize}[2]{\resizebox{#1}{!}{$#2$}}%
\newcommand{\BlueText}[1]{\textcolor{blue}{#1}}
\newcommand{\remove}[1]{ }

\makeatletter
\setlength{\@fptop}{0pt}
\makeatother

\newcommand{\claudio}[1]{\textcolor{blue}{{#1}}}
\newcommand{\henry}[1]{\textcolor{red}{{#1}}}

\newcommand{\R}{\mathbb{R}}
\newcommand{\Z}{\mathbb{Z}}

\title{Architecture Design for Human-Driven Systems}

\author{
\IEEEauthorblockN{Mahyar T. Moghaddam\thanks{mtmo@mmmi.sdu.dk}\IEEEauthorrefmark{1}, {Moamin B. Abughazala}\thanks{moamin.abughazala@graduate.univaq.it}\IEEEauthorrefmark{2}, {Vittorio Cortellessa}\thanks{vittorio.cortellessa@univaq.it}\IEEEauthorrefmark{2}, {Antinisca Di Marco}\thanks{antinisca.dimarco@univaq.it}\IEEEauthorrefmark{2}, {Henry Muccini}\thanks{henry.muccini@univaq.it}\IEEEauthorrefmark{2},\\ {Fabrizio Rossi}\thanks{fabrizio.rossi@univaq.it}\IEEEauthorrefmark{2}, {Karthik Vaidhyanathan}\thanks{karthik.vaidhyanathan@univaq.it}\IEEEauthorrefmark{2}}
\IEEEauthorblockA{\IEEEauthorrefmark{1}\textit{MMMI Institute}, \textit{University of Southern Denmark},
Odense, Denmark \\}
\IEEEauthorblockA{\IEEEauthorrefmark{2}\textit{DISIM Department},
\textit{University of L'Aquila},
L'Aquila, Italy}
}

\maketitle

\begin{abstract}
This paper highlights humans' social and mobility behaviors' role in the continuous engineering of sustainable socio-technical systems. Our approach relates the humans' characteristics and intentions with the system's goals, and models such interaction. Such a modeling approach aligns the architectural design and associated quality of service (QoS) with humans' quality of experience (QoE).
We design a simulation environment that combines agent-based social simulation (ABSS) with architectural models generated through a model-driven engineering approach. Our modeling approach facilitates choosing the best architectural model and system configuration to enhance both the humans' and system's sustainability.
We apply our approach to the Uffizi Galleries crowd management system. Taking advantage of real data, we model different scenarios that impact QoE.
We then assess various architectural models with different SW/HW configurations to propose the optimal model based on different scenarios concerning QoS-QoE requirements. 
\end{abstract}


\section{Introduction}
Architectural design decisions are historically driven by technical reasoning and concerns. 
More recently, other aspects such as business value 
and sustainability. 
The growing importance of sustainable development is highlighting additional concerns regarding social, individual, economics, and environmental interlinked with technical dimensions \cite{becker2015sustainability}. Literature on sustainable software development generally focuses on environmental aspects; still, little attention is dedicated to {\em social and individual} human sustainability.
In this paper, we propose a {\em human-oriented architecture design approach} for socio-technical systems. We analyze how the understanding of human behavior may drive the selection of different and alternative architectural models \cite{muccini2018iot} and configurations, with the objective of minimizing energy consumption.
In order to do so, we present a comprehensive approach comprising: {\em i)} an agent-based modeling (ABM) to model humans behavior,
{\em ii)} an ABM and model-driven engineering integration engine,
{\em iii)} a simulation environment (fed by real data) that simulates human behavior with respect to Internet of Tings (IoT) resources,
{\em iv)} the application on the Uffizi Galleries crowd management system.

The paper is organized as follows. Background is presented in Section II. Our agent-based and architectural modeling methodologies are presented in Section III. The method is applied to a real case study in Section IV, and the conclusions are finally drawn in Section V.

\section{Background}

In {\em agent-based modeling (ABM)}, the environment and what is included in it are modeled as agents \cite{dugdale2019agent}. Each agent has a set of characteristics and behaviors. 
For simulation of human behavior, agent-based social simulations (ABSS) are generally used.
In ABSS, an agent is defined as an autonomous software entity that can act upon and perceive its environment. 
When agents are put together, they form an artificial society, each perceiving, moving, performing actions, communicating, and transforming the local environment, much like human beings in real society.
In this paper, we use PedSim \footnote{https://www.pedsim.net/} 
to simulate IoT environment and moving agents.
{\em Modeling the IoT system's architecture} implies considering the IoT components, their interactions, the underlying hardware configurations of the different IoT components, as well as constraints from the environment in which these IoT components will be deployed. To this end, we use our CAPS modeling framework\footnote{http://caps.disim.univaq.it/}~\cite{caps} to provide a multi-view (software, hardware, physical space) approach.
In this work, we use CupCarbon, a state-of-the-art smart city IoT simulator\footnote{http://cupcarbon.com/}, 
widely used in research, especially for energy and data simulation of IoT systems.


\section{Methodology}

This section presents the methodology that links ABM and ABSS with IoT architectural modeling and simulation, thereby allowing architects to analyze the models and design an optimal architecture with respect to QoS and QoE requirements.
Our methodology consists of four stages:
{\em i) Agent Modeling and Simulation} stage that deals with the modeling and simulation of different agents using the ABM and ABSS approaches. The agent-based model for IoT socio-technical systems consists of four classes of agents: humans, cyber elements, physical space, and IoT resources which all are part of the environment class.
{\em ii) Agent-IoT composition} stage receives the results of the agent-based simulations to generate the data required for the IoT simulation. This step consists of a single process, namely, Agent IoT Data Composition (AIDC).
{\em iii) IoT modeling and simulation} stage involves modeling and simulation of different architectural models and configurations based on the data received from the agent-IoT composition step.
{\em iv) Analysis} stage processes the IoT simulation results by taking into account the QoS/QoE goals (which can be specified by stakeholders). It then identifies the optimal architectural model and configuration for a given agent behavior. It achieves this with the help of a utility metric, namely trade-off score, $t_{s} = w_s*Q_s + w_e*Q_e$; where $w_s$ and $w_e$ represents the weights given to QoS and QoE goals respectively and $Q_s$ and $Q_e$ are piece-wise functions that captures the satisfaction of QoS and QoE goals respectively.

\section{Results}

We applied our approach on {\em Uffizi Gallery} crowed management, the most visited museums in Florence, Italy \cite{dugdale2020human}.
In our agent-based model (ABM) of the Uffizi galleries, we set the simulation parameters either by real data gathered or according to the literature. We considered two scenarios. In the {\bf \em first scenario} we used the Uffizi historical data to assign different characteristics to human agents regarding age, gender, origin, and physical condition. In the {\bf \em second scenario} we modeled social attachment and grouping in Uffizi to assess its impact on QoE and QoS.
Getting input from the ABSS, we modeled the IoT architecture of the Uffizi case study using the CAPS modeling framework. 
In the Uffizi museum, IoT devices need to be deployed in the entrance, exits, and corridors to monitor the movement of humans into, outside of, and within the museum. The IoT device's choice can impact the overall QoS and QoE offered by the system.
To define the configuration of the IoT components in the models, we use the concept of modes provided by CAPS. Every IoT component operates in two modes, {\em i)} normal mode when the sensor reads/sends data at a lower frequency; {\em ii)} critical mode where the sensor reads/sensor at a higher frequency due to some critical condition (high queuing). However, the choice of the frequency in either mode can impact the QoS and QoE. We performed 36 simulations ($6$ models $\times$ $3$ configurations $\times$ $2$ scenarios). Each simulation was performed for a total time of 2 hours in a Windows-based desktop machine running on an Intel i7, 2.6-3.2 GHz processor with 16 Gb RAM.
We analyze the results to identify the optimal architectural model and configuration for the two scenarios modeled by ABSS.

Regarding {\bf \em Scenario 1} (Congestion), we calculate the trade-off score ($t_s$) by giving more weight to QoE.
We set the goals such that we want the system to consume not more than 100 joules and the server to capture the movement of at least 1200 people every 15 minutes. Our simulations show that using 1 counter in the entrance, 3 counters in exit, and a mix of 5 cameras and 4 RFID in corridors provides the highest $t{s}$. In the mentioned configuration, in normal and critical modes, cameras and RFID readers send packets with a frequency of 30 and 10 per second, and people counters send them by a frequency of 10 and 1, respectively. This configuration gives more preference to QoE without compromising too much on the QoS.
In {\bf \em scenario 2}, we calculated the trade-off score, $t_{s}$ for each of the model configuration pairs, with the goal of consuming no more than 150 joules and the server to capture the movement of at least 800 people every 15 minutes. We observed that as opposed to Scenario 1, using 1 QR Reader in the entrance, 3 Counters in exit and 9 RFID in corridors provides the highest $t{s}$. 
In the mentioned configuration, in normal and critical modes, QR Reader sends packets with a frequency of 20 and 5, RFID sends them with a frequency of 40 and 20, and counters send them with a frequency of 20 and 5.
The above evaluations indicate how our approach could allow architects to model the expected human behavior and select the appropriate architectural models and configurations to optimize the overall QoS/QoE (or a combination) offered by the system. Moreover, our approach provides architects with a set of models and configurations to handle different human behavioral scenarios.


\section{Conclusion}

In this work, we demonstrated that using a human-oriented approach can allow software architects to better design socio-technical systems. It achieves this with the help of a model-driven approach which enables architects to model the expected behaviors of humans, the architecture of the IoT system and further provides mechanisms to simulate and perform trade-off analysis of different design alternatives with respect to QoE and QoS requirements. Our evaluation on a real case study shows that our approach can allow architects to select optimal architectural models and configurations by considering the expected human behavior.
We plan to evolve this approach into a tool that can be used by software architects to better design socio-technical systems.

\section*{Acknowledgment}
We acknowledge the support given by the Uffizi Galleries and its director Dr. Eike Schmidt. This research is supported by the VASARI PON R\&I 2014-2020 and FSC project.

\bibliographystyle{unsrt}

\bibliography{bib}

\end{document}